\documentclass[prl,twocolumn,amsmath,amssymb,floatfix,superscriptaddress,nofootinbib,preprintnumbers, longbibliography]{revtex4-2}
\pdfoutput=1

\usepackage{amssymb}
\usepackage{amsmath}

\usepackage{graphicx}
\usepackage{graphics}
\usepackage{dcolumn}
\usepackage[dvipsnames]{xcolor}
\usepackage{fancyhdr} 
\usepackage{graphicx}
\usepackage{float}
\usepackage{multirow}

\usepackage{subfig}
\usepackage[colorlinks=true,linktocpage=true,urlcolor=blue,citecolor=blue,linkcolor=blue]{hyperref}

\usepackage[utf8]{inputenc}

\usepackage{dsfont}

\usepackage[justification=centerlast, format=plain, labelfont=bf]{caption}

\makeatletter
\renewcommand\tableofcontents{%
    \@starttoc{toc}%
}
\makeatother

\DeclareCaptionJustification{justified}{\leftskip=0pt \rightskip=0pt \parfillskip=0pt plus 1fil}

    \newcommand{\be}{\begin{equation}}
  \newcommand{\ee}{\end{equation}}
    \newcommand{\ba}{\begin{align}}
  \newcommand{\ea}{\end{align}}

\newcommand{\Msun}{M_{\odot}}
\newcommand{\Mpcinv}{ {\rm Mpc}^{-1} }

\newcommand{\MUV}{ M_{\rm UV} }

\makeatletter
\def\doauthor#1#2#3{%
  \ignorespaces#1\unskip
  \begingroup
   #3%
  \@if@empty{#2}{\@listcomma\endgroup{}{}}{\endgroup{\comma@space}{}\frontmatter@footnote{#2}}%
  \space \@listand
}%
\makeatother

\makeatletter
\def\@ssect@ltx#1#2#3#4#5#6[#7]#8{%
  \def\H@svsec{\phantomsection}%
  \@tempskipa #5\relax
  \@ifdim{\@tempskipa>\z@}{%
    \begingroup
      \interlinepenalty \@M
      #6{%
       \@ifundefined{@hangfroms@#1}{\@hang@froms}{\csname @hangfroms@#1\endcsname}%
       {\hskip#3\relax\H@svsec}{#8}%
      }%
      \@@par
    \endgroup
    \@ifundefined{#1smark}{\@gobble}{\csname #1smark\endcsname}{#7}%
  }{%
    \def\@svsechd{%
      #6{%
       \@ifundefined{@runin@tos@#1}{\@runin@tos}{\csname @runin@tos@#1\endcsname}%
       {\hskip#3\relax\H@svsec}{#8}%
      }%
      \@ifundefined{#1smark}{\@gobble}{\csname #1smark\endcsname}{#7}%
      \addcontentsline{toc}{#1}{\protect\numberline{}#8}%
    }%
  }%
  \@xsect{#5}%
}%
\makeatother

\begin{document}

\title{\vspace*{-0.15cm} Insights from HST into Ultra-Massive Galaxies and Early-Universe Cosmology}

\author{Nashwan Sabti$^{\mathds{S},}$}
\affiliation{William H. Miller III Department of Physics and Astronomy, 3400 N. Charles St., Baltimore, MD 21218, USA}
\author{Julian B.~Mu\~{n}oz$^{\mathds{M},}$}
\affiliation{Department of Astronomy, The University of Texas at Austin, 2515 Speedway, Stop C1400, Austin, TX 78712, USA}
\affiliation{Center for Astrophysics $|$ Harvard \& Smithsonian, 60 Garden St., Cambridge, MA 02138, USA}
\author{Marc Kamionkowski$^{\mathds{K},}$}
\affiliation{William H. Miller III Department of Physics and Astronomy, 3400 N. Charles St., Baltimore, MD 21218, USA}

\def\thefootnote{$\mathds{S}$\hspace{0.7pt}}\footnotetext{\href{mailto:nsabti1@jhu.edu}{nsabti1@jhu.edu}}
\def\thefootnote{$\mathds{M}$\hspace{-0.9pt}}\footnotetext{\href{mailto:julianmunoz@austin.utexas.edu}{julianmunoz@austin.utexas.edu}}
\def\thefootnote{$\mathds{K}$}\footnotetext{\href{mailto:kamion@jhu.edu}{kamion@jhu.edu}}
\setcounter{footnote}{0}
\def\thefootnote{\arabic{footnote}}

\begin{abstract} 
The early-science observations made by the James Webb Space Telescope (JWST) have revealed an excess of ultra-massive galaxy candidates that appear to challenge the standard cosmological model ($\Lambda$CDM). Here, we argue that any modifications to $\Lambda$CDM that can produce such ultra-massive galaxies in the early Universe would also affect the UV galaxy luminosity function (UV LF) inferred from the Hubble Space Telescope (HST). The UV LF covers the same redshifts ($z\approx 7-10$) and host-halo masses $(M_\mathrm{h}\approx 10^{10}-10^{12}\, M_\odot$) as the JWST candidates, but tracks star-formation rate rather than stellar mass. 
We consider beyond-$\Lambda$CDM power-spectrum enhancements and show that any departure large enough to reproduce the abundance of ultra-massive JWST candidates is in conflict with the HST data. 
Our analysis, therefore, severely disfavors a cosmological explanation for the JWST abundance problem. Looking ahead, we determine the maximum allowable stellar-mass function and provide projections for the high-$z$ UV LF given our constraints on cosmology from current HST data.
\end{abstract}

\maketitle

\textit{Introduction.\ ---} 
The study of the earliest generations of galaxies offers a unique opportunity to understand the Universe during the epochs of cosmic dawn and reionization at redshifts $z\sim 5-30$. This era represents a largely uncharted territory in the timeline of the Universe, occupying the gap between cosmic-microwave-background (CMB) decoupling ($z\sim 10^3$) and the local Universe ($z\sim 0$), thus making its investigation a crucial step towards a complete understanding of cosmology.

The early-science observations made by the James Webb Space Telescope (JWST) have unveiled a multitude of galaxy candidates in this era~\cite{Finkelstein_CEERS_I1,Treu:2022iti,Castellano_GLASS_hiz,Harikane_UVLFs,Naidu_UVLF_2022,PerezGonzalez_CEERS, Papovich_CEERS_IV,Santini_GLASS_Mstar,Ilie:2023zfv}. While only a subset of these candidates have been spectroscopically confirmed so far~\cite{Arrabal_CEERS_spectra,Arrabal_CEERS_spectra2,Robertson:2022gdk,CurtisLake_JWST_spectra, Fujimoto:2023orx,2022arXiv221015699Williams, 2022arXiv221015639Borsani, 2023arXiv230107072Tang,Kocevski_CEERS_spectra_AGN}, the findings from JWST provide an intriguing glimpse into a highly-active Universe during this formative era, with ultra-massive galaxies possibly playing a significant role. The longer-wavelength capabilities of JWST exceed those of the Hubble Space Telescope (HST), enabling measurements of rest-frame visible light in addition to the UV. This feature makes it possible to track the total stellar mass in a galaxy, in addition to its star-formation rate, providing valuable insights into astrophysics and cosmology at high redshifts.

In a recent study, early-release photometric data from JWST were used to estimate the stellar mass of massive galaxy candidates at redshifts $z\approx 7 - 10$~\cite{Labbe_2023}. 
Interestingly, two of the reported objects were claimed to exhibit stellar masses significantly larger than what would be expected within the standard cosmological model ($\Lambda$CDM). If confirmed (though we note Ref.~\cite{Fujimoto:2023orx} reports a lower stellar mass for one of these objects using spectroscopy), this finding would not only challenge our models of galaxy formation, but also of cosmology. Indeed, assuming that galaxies form within dark-matter halos, Refs.~\cite{Boylan-Kolchin:2022kae,Lovell:2022bhx} presented a compelling argument that $\Lambda$CDM predicts an insufficient number of halos in the observed volume to host these galaxies, even if most baryons in the galaxies were converted into stars. Consequently, this poses a significant {\it abundance problem} for the $\Lambda$CDM paradigm.

This discrepancy raises questions about the validity of our current astrophysical and cosmological models. As a result, there has been a surge in theoretical work aimed at modifying $\Lambda$CDM to explain this abundance issue~\cite{Liu:2022bvr,Hutsi:2022fzw, Menci:2022wia, Biagetti:2022ode,Jiao:2023wcn,Parashari:2023cui}. The goal of such modifications is to generate additional galaxies at $z= 7-10$ by enhancing structure formation during this period, while preserving the successful $\Lambda$CDM predictions at the CMB and lower-$z$ epochs.

Here we point out that UV luminosity function (UV LF) data obtained with HST have already probed the same range of redshifts and distance scales as the galaxies from Ref.~\cite{Labbe_2023}. Specifically, using these data, we demonstrate that generic enhancements of power that could explain the observations from Ref.~\cite{Labbe_2023} are ruled out, hence disfavoring a cosmological solution to the JWST abundance problem. Finally, in anticipation of future surveys, we determine the maximum permissible cosmological increase in the stellar-mass function at these redshifts and provide predictions for the high-$z$ UV LF.

Throughout this paper, we will assume a flat cosmology and, unless otherwise stated, fix the
cosmological parameters to the Planck 2018 best-fits~\cite{Planck:2018vyg}: $h = 0.6727$, $\omega_\mathrm{b} = 0.02236$, $\omega_\mathrm{cdm} = 0.1202$, $\ln\left(10^{10}A_\mathrm{s}\right) = 3.045$, $n_\mathrm{s} = 0.9649$, and $\tau_\mathrm{reio} = 0.0544$.\\

\newpage

\textit{Cosmology from HST UV LFs.\ ---} The deep-field surveys conducted by the Hubble Space Telescope have proven to be invaluable in advancing our understanding of early-Universe astrophysics~\cite{Mason:2015cna,Tacchella:2018qny,Yung_2018,Gillet:2019fjd}. These surveys contain galaxies dating back as far as redshift $z \sim 10$, deep into the epoch of cosmic reionization~\cite{Bouwens:2014fua,Finkelstein_2015,Atek:2015axa,Livermore:2016mbs,Bouwens_2017asdasd,Mehta_2017,Ishigaki_2018,Oesch_2018,Atek:2018nsc, Rojas_Ruiz_2020, Bouwens_2021}. Along with their significant role in astrophysics, the UV luminosity functions derived from these same data provide us with new insights into cosmology in a relatively unexplored range of times and scales. Indeed, since early galaxies reside in dark-matter halos, their observations enable us to indirectly track the cosmological halo mass function. Consequently, the HST UV LFs have been used to set constraints within various cosmological scenarios, including those featuring a suppression of power from dark matter~\cite{Bozek:2014uqa, Schultz:2014eia,  Corasaniti:2016epp,Menci:2017nsr,Rudakovskyi:2021jyf} and enhancements from inflationary physics~\cite{Sabti:2020ser, Chevallard:2014sxa,Yoshiura:2020soa}, as well as to provide model-agnostic measurements of the matter power spectrum~\cite{Sabti:2021unj}.

\begin{figure}[t!]
    \centering
\includegraphics[width=\linewidth]{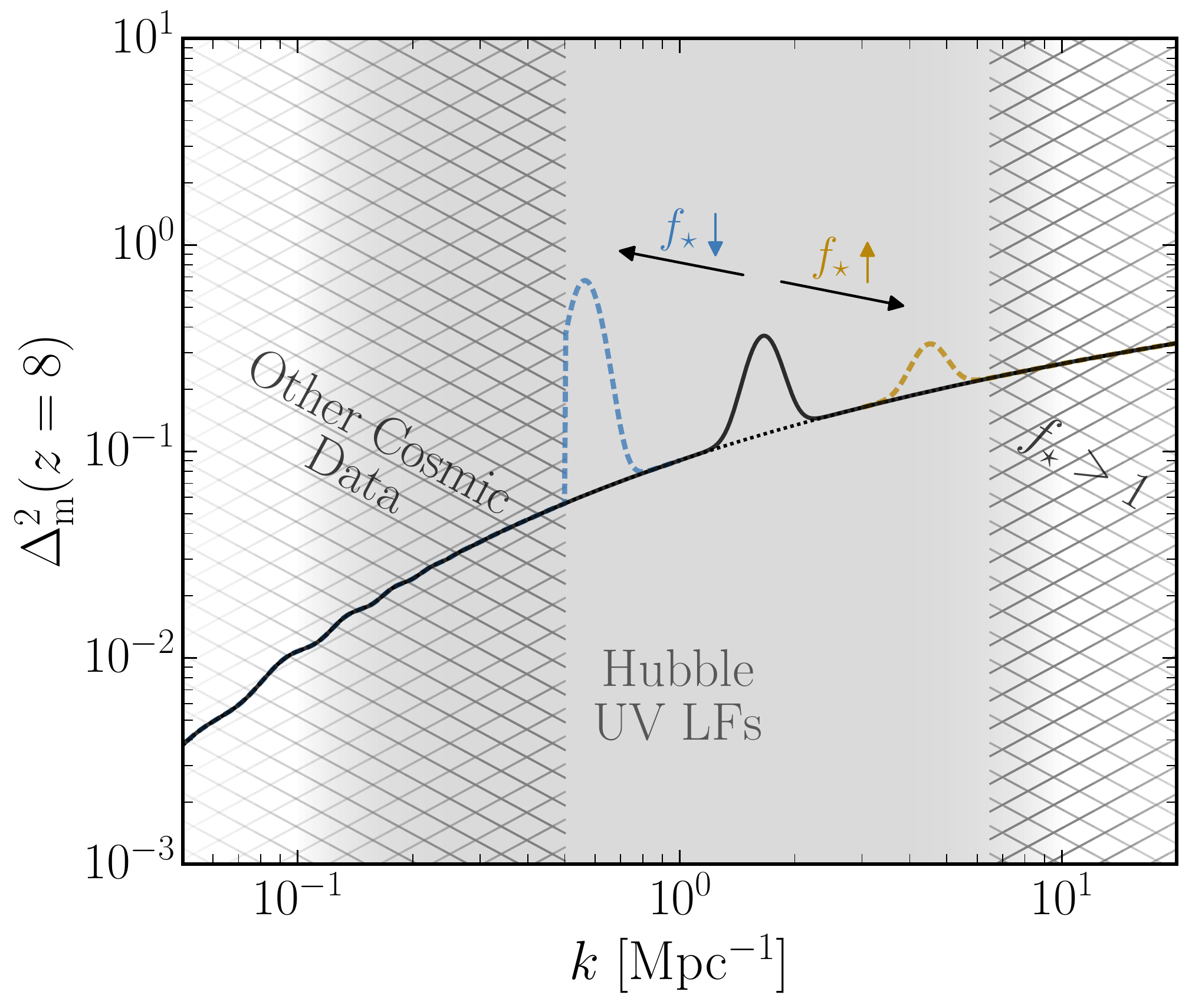}
    \caption{The dimensionless matter power spectrum, linearly extrapolated to $z=8$. The $\Lambda$CDM prediction (black dotted) is not sufficient to explain the observed abundance of massive JWST galaxy candidates. We show the power enhancement (i.e., the bump described in Eq.~\eqref{eq:bump}) that is necessary to address this abundance problem. The location and amplitude of the enhancement depend on the assumed star-formation efficiency $f_\star$. For instance, a lower $f_\star$ shifts the bump towards smaller $k$, as more massive halos are needed, and requires a larger amplitude due to the rarity of such halos. The gray shaded band represents the range of scales probed by the Hubble UV LFs, while the hatched areas indicate the scales at which other cosmological probes are sensitive, or when $f_\star > 1$ is required.}
\label{fig:pk_with_bumps}
\end{figure}

We illustrate the reach of the HST UV LFs in Fig.~\ref{fig:pk_with_bumps} through the dimensionless matter power spectrum (the two-point correlation function in Fourier-space). The HST data probe redshifts $z \leq 10$ and length scales $0.1\, \Mpcinv \lesssim k \lesssim 10\, \Mpcinv$~\cite{Sabti:2021unj}, covering the entire range where proposed solutions to the JWST abundance problem alter cosmology. As such, the UV LFs can shed light on whether there is an excess of power (and, hence, galaxies) at high redshifts. We will leverage the public code {\tt GALLUMI}\footnote{\url{https://github.com/NNSSA/GALLUMI_public}}\,\cite{Sabti:2021xvh} to analyze the HST observations. {\tt GALLUMI} provides a pipeline for modeling the UV LFs and comparing them with data through an MCMC analysis. This enables us to simultaneously constrain astrophysics and cosmology, and thus determine how much power
enhancement is allowed by the HST data.\\

\textit{Confronting JWST Massive Galaxies.\ ---}
By assuming that each dark-matter halo at high $z$ contains a single central galaxy, the expected number of galaxies with stellar mass larger than some observational threshold $M_\star^{\rm obs}$ can be obtained as~\cite{Boylan-Kolchin:2022kae}:
\begin{align}
\label{eq:Ngal}
N_{\rm gal}(M_\star \geq M_\star^{\rm obs}) &= \mathrm{Vol} \times \int_{M_\mathrm{h}^{\rm cut}}^\infty \mathrm{d}M_\mathrm{h} \dfrac{\mathrm{d}n}{\mathrm{d}M_\mathrm{h}}(z_\mathrm{obs}, M_\mathrm{h})\ ,
\end{align}
where Vol $=\Omega(r_{\rm max}^3 - r_{\rm min}^3)/3$ is the comoving survey volume, determined by the surveyed redshift range (from $r(z_\mathrm{min})$ to $r(z_\mathrm{max})$) and the survey angular coverage $\Omega$, and $\mathrm{d}n/\mathrm{d}M_\mathrm{h}$ is the halo mass function\footnote{We will use the Sheth-Tormen~\cite{Sheth:2001dp} mass function as in Ref.~\cite{Sabti:2021xvh}, but with a fit calibrated to high-redshift N-body simulations from Ref.~\cite{Schneider:2020xmf}.}. 
The halo-galaxy connection enters through the relation between the cut-off scales in stellar/halo mass $M_{\star/\mathrm{h}}^{\rm obs/cut}$. We will follow a form of abundance matching, where we assume that a fraction $f_\star \leq 1$ of the baryons inside a halo get converted into stars\footnote{Note that feedback effects, which typically reduce the star-formation efficiency $f_\star$ in heavy halos~\cite{Wechsler:2018pic}, are not considered here, making this a conservative approach.}. As a result, we can link halo masses to stellar masses  simply as $M_\mathrm{h} = M_\star/(f_\mathrm{b} f_\star)$ for a given $f_\star$, where we keep the baryon fraction fixed to $f_\mathrm{b} = 0.157$~\cite{Planck:2018vyg}, as done in Refs.~\cite{Boylan-Kolchin:2022kae,Behroozi:2016mne}. 

Following Ref.~\cite{Boylan-Kolchin:2022kae}\footnote{See a related analysis in Ref.~\cite{Lovell:2022bhx} in the language of extreme-value statistics.}, we will use the brightest galaxy candidate in each of the two redshift bins reported in~\cite{Labbe_2023}\footnote{Ultra-massive high-$z$ candidates have also been reported in Refs.~\cite{Tachella_Mstar_HST,Rodighiero_JWST_Mstar}.} that are in most tension with $\Lambda$CDM, which we summarize in Tab.~\ref{tab:summary}. The CEERS survey where these candidates were found covers an area of $\Omega = 38\ \mathrm{arcmin^2}$~\cite{Finkelstein_CEERS_I1}, which in $\Lambda$CDM encompasses an effective volume of $\mathrm{Vol} \approx 10^5\ \mathrm{Mpc}^3$ in both redshift bins. To explore the impact of the star-formation efficiency $f_\star$, we will consider three fiducial choices: $f_\star = 0.1$, the peak of the star-to-halo mass ratio~\cite{Wechsler:2018pic,Stefanon_2021_Spitzer_fstar}, an enhanced star-efficiency scenario with $f_\star = 0.3$, and an astrophysically implausible situation with $f_\star=0.5$, which we use as a conservative limit. By employing Eq.~\eqref{eq:Ngal}, we can then determine the average number of galaxies expected in a survey. To put this number into perspective, within $\Lambda$CDM we predict $N_\mathrm{gal}^\mathrm{\Lambda CDM} = 2.3\times10^{-2}\ll 1$ for $f_\star = 0.5$. That is, even in this astrophysically unrealistic scenario $\Lambda$CDM predicts an abundance that is significantly smaller than what is reported in Ref.~\cite{Labbe_2023}. To accurately model the Poissonian nature of the observations, we assign a Poisson likelihood to each galaxy in the sample under consideration:
\begin{align}
\label{eq:JWST_lkl}
\ln\mathcal{L} = \sum_{\mathrm{Galaxies}}\left[-N_\mathrm{gal}^\mathrm{pred} + N_\mathrm{gal}^\mathrm{obs}\ln\left(N_\mathrm{gal}^\mathrm{pred}\right) - \ln\left(N_\mathrm{gal}^\mathrm{obs}!\right)\right]\ ,
\end{align} 
where $N_\mathrm{gal}^\mathrm{obs} = 1$ is the JWST observation in each of the two redshift bins in Tab.~\ref{tab:summary}, and $N_\mathrm{gal}^\mathrm{pred}$ is the prediction from Eq.~\eqref{eq:Ngal}, which depends on the underlying cosmology\footnote{While we do not account for cosmic variance in Eq.~\eqref{eq:JWST_lkl}, the Poisson errors dominate the total error budget~\cite{Trenti:2007dh}.}. We can now study what changes to the latter are required to explain the abundance of JWST candidates.

\begin{table}[t!]
\centering
{\def\arraystretch{1.35}
\begin{tabular}{c|c|c|c}
\hline\hline
\textbf{Galaxy ID} & $\boldsymbol{\log_{10}\left(M_\star/M_\odot\right)}$ & $\boldsymbol{z_\mathrm{obs}}$ & $\boldsymbol{z_\mathrm{min} - z_\mathrm{max}}$\\
\hline\hline
38094 & 10.89 & 7.48 & $7 - 8.5$  \\
\hline
35300 & 10.40 & 9.08 & $8.5 - 10$ \\
\hline\hline
\end{tabular}
}
\caption{Summary of the properties of two massive JWST galaxy candidates identified in~\cite{Labbe_2023}.}
\label{tab:summary}
\end{table}

The most straightforward cosmological solution to the abundance problem is to amplify the matter power spectrum, thereby increasing the halo mass function $\mathrm{d}n/\mathrm{d}M_\mathrm{h}$ in Eq.~\eqref{eq:Ngal}. We thus consider an effective enhancement of the matter power spectrum as $P(k) = P_{\Lambda \rm CDM}(k) \left[1 + A(k)\right]$,
where the amplitude boost is given by:
\begin{align}
\label{eq:bump}
A(k) = A_\star \exp\left[-\frac{1}{2}\left(\frac{\ln(k) - \ln(k_\star)}{\sigma_\star}\right)^2\right]\ ,
\end{align}
with amplitude $A_\star$, mean $\ln(k_\star)$, and a (log) width $\sigma_\star$. This effective model captures a generic boost of power during cosmic dawn and reionization that could have been missed by observations at both lower and higher redshifts. To avoid violating CMB lensing measurements, whose $z$ kernel covers part of the redshift range under consideration~\cite{Lewis:2006fu}, we will set the enhancement equal to zero for wavenumbers $k\leq 0.5\ \Mpcinv$~\cite{Chabanier:2019eai}. We relax this assumption in the Supplementary Material (which includes Refs.~\cite{Reed:2006rw, Meurer:1999jj, Bouwens:2011yy, Bouwens:2013hxa, Overzier:2010aa, Casey:2014cqa}), where we show that the HST UV LFs also disfavor a broad-band enhancement as a viable solution to the abundance problem.

In Fig.~\ref{fig:pk_with_bumps}, we illustrate the boosts in power generated by Eq.~\eqref{eq:bump}. In all cases, the JWST data require a fairly large bump on top of $\Lambda$CDM to reproduce the abundance of the JWST galaxies, with a location and amplitude that depend on the assumed star-formation efficiency $f_\star$. For low values of $f_\star$, a boost with larger amplitude $A_\star$ and smaller mean $k_\star$ is required. This is because a decrease in $f_\star$ results in an increase in the host halo mass of these galaxies, and heavier halos are rarer and produced from lower-$k$ fluctuations. Conversely, an increase in $f_\star$ leads to the \emph{possibility} that the enhancement could now be located at smaller scales (higher $k$) as well. The Hubble UV LFs can probe the relevant range of scales for $f_\star\leq 1$, as demonstrated in Fig.~\ref{fig:pk_with_bumps}. Our next step then is to test whether a cosmological solution to the JWST abundance problem conforms to HST UV LF data. We will do so in two steps.

In the first part, which we label \emph{JWST}, we will use the likelihood in Eq.~\eqref{eq:JWST_lkl} to determine the power enhancement necessary to account for the abundance of the JWST galaxies reported in Ref.~\cite{Labbe_2023}. We calculate the modified halo mass function given the boost in Eq.~\eqref{eq:bump} and obtain the predicted number of galaxies $N_\mathrm{gal}^\mathrm{pred}$ with Eq.~\eqref{eq:Ngal}. We implement this in the MCMC sampler \texttt{MontePython}~\cite{Brinckmann:2018cvx}, where we vary the amplitude and mean of the boost $A(k)$, while fixing the width to $\sigma_\star = 0.1$. We test this latter assumption, along with others, in the Supplementary Material, finding no discernible difference.

In the second analysis, labeled \emph{HST}, we will use the HST UV LFs to {\it constrain} the same boosts in power. We employ the \texttt{GALLUMI}~\cite{Sabti:2021xvh} pipeline for this purpose, and modify it to include the same $A(k)$ parameterization. We consider only the Hubble Legacy Fields data from Ref.~\cite{Bouwens_2021_UVLFs} within the redshift range\footnote{We choose to cover roughly the same redshifts as the JWST candidates from Ref.~\cite{Labbe_2023}, though we emphasize that UV LF data is also available at lower redshifts, which would further strengthen our constraints.} $z=6-10$. While we keep the base cosmological parameters fixed, we allow for variations in the astrophysical parameters together with the amplitude and mean of the power-spectrum enhancement. Further details on the \texttt{GALLUMI} pipeline, including the fiducial astrophysical model and statistical approach used, can be found in Ref.~\cite{Sabti:2021xvh}.

\begin{figure}[t!]
    \centering
    \includegraphics[width=\linewidth]{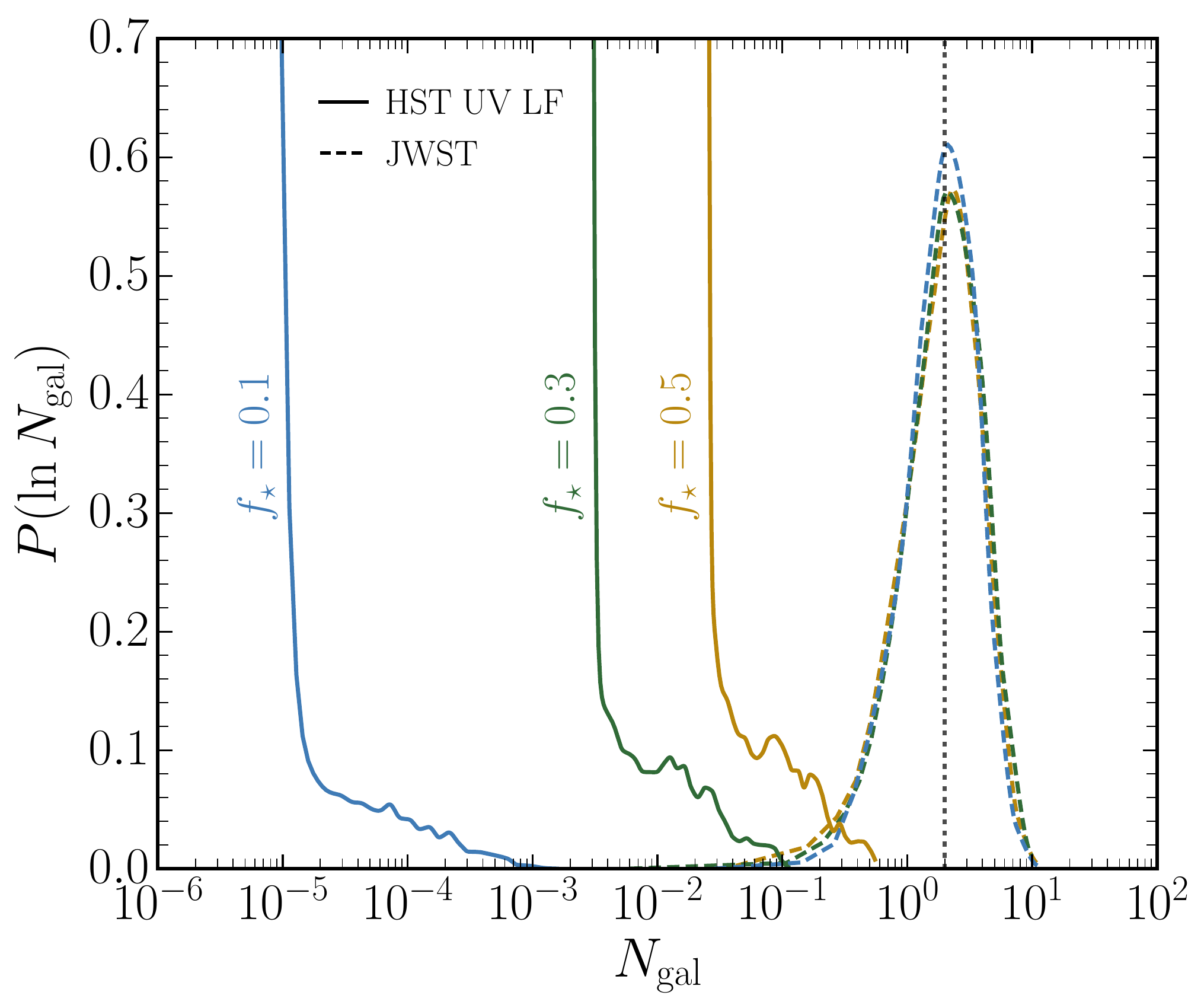}
    \caption{Probability distributions of the average number of galaxies, calculated using Eq.~\eqref{eq:Ngal} for the two brightest candidates in the JWST sample from~\cite{Labbe_2023}. The solid lines correspond to the results obtained using HST UV LF data with the \texttt{GALLUMI} pipeline, while the dashed lines are the [Poisson—cf., Eq.~\eqref{eq:JWST_lkl}] probability distributions inferred from the two observed JWST galaxies. It is clear that the HST posteriors do not reach the observed value of $N_{\rm gal}=2$ (black dotted line). We consider three different values of the star-formation efficiency $f_\star$, at or above the typical peak height of $f_\star=0.1$, to explore the impact of astrophysics.
    }
    \label{fig:Ngal_distributions}
\end{figure}

We can now compare the results of our JWST and HST analyses. After each run, we obtain chains for $A_\star$ and $k_\star$ (see the Supplementary Material), from which we compute $N_\mathrm{gal}$ as a derived parameter using Eq.~(\ref{eq:Ngal}). We show the posterior distributions for $\ln(N_\mathrm{gal})$ in Fig.~\ref{fig:Ngal_distributions} for our three choices of $f_\star$. 
The JWST posteriors all peak around $N_{\rm gal}=2$ (corresponding to the two galaxies considered in our sample), showing that our model for $A(k)$ has enough freedom to explain the JWST data. That is, by adding sufficient power, we can give rise to the necessary number of heavy halos and therefore galaxies. In contrast, the HST UV LF posteriors show that only a modest enhancement of the number of galaxies beyond the $\Lambda$CDM prediction -- which corresponds to the lowest $N_{\rm gal}$ value for each $f_\star$ in Fig.~\ref{fig:Ngal_distributions} -- is allowed. In particular, the HST data does not allow $N_{\rm gal} = 2$ for any $f_\star \leq 0.5$, and even for the extreme case of $f_\star=1$ we find it in mild ($P(N_{\rm gal}\geq 2) < 1.2 \%$) tension. These findings broadly hold for $N_{\rm gal} = 1$ as well. We therefore conclude that any power enhancement induced to explain the JWST abundance problem is in significant tension with HST UV LF data at the same redshifts and distance scales.

While we have only parameterized cosmological departures through Eq.~\eqref{eq:bump}, our constraints could be broadly applicable to models that enhance the halo abundance at $z=7-10$ through other means (e.g., non-Gaussianities~\cite{Sabti:2020ser,Biagetti:2022ode}, inflationary features~\cite{Kosowsky:1995aa,Kamionkowski:1999vp,Seleim:2020eij}, or modifications to the expansion rate~\cite{Menci:2022wia}), as the constraints from the HST UV LFs are directly linked to the abundance of halos that could host ultra-massive galaxies (though a dedicated analysis may be required). We also note that not every galaxy need be UV-bright (e.g., the ones from Ref.~\cite{Labbe_2023}), given the stochasticity we expect in the $M_\mathrm{h}$-$\MUV$ connection~\cite{Ren_2019_scatter}. Nevertheless, both our model and the {\it Subaru} clustering measurements from Ref.~\cite{Harikane_Goldrush_2021} ensure that the halos with $M_\mathrm{h}\sim 10^{12}\,\Msun$ host galaxies that are UV-bright on average at the relevant redshifts, and thus that the UV LF tracks the abundance of these halos. The strength of the HST UV LFs is that they are {\it contemporaneous} to the JWST galaxies of Ref.~\cite{Labbe_2023}, and, as such, are probing the same underlying halo mass function.  Our analysis, of course, does not exclude astrophysical explanations for the high $M_\star$ measurements, including Pop III star formation~\cite{Steinhardt_PopIII}, a local baryonic overabundance~\cite{Chen_ultramassive}, or (active-galactic-nuclei) line contamination~\cite{Endsley_2023}.\\

\begin{figure}[t!]
    \centering
    \includegraphics[width=\linewidth]{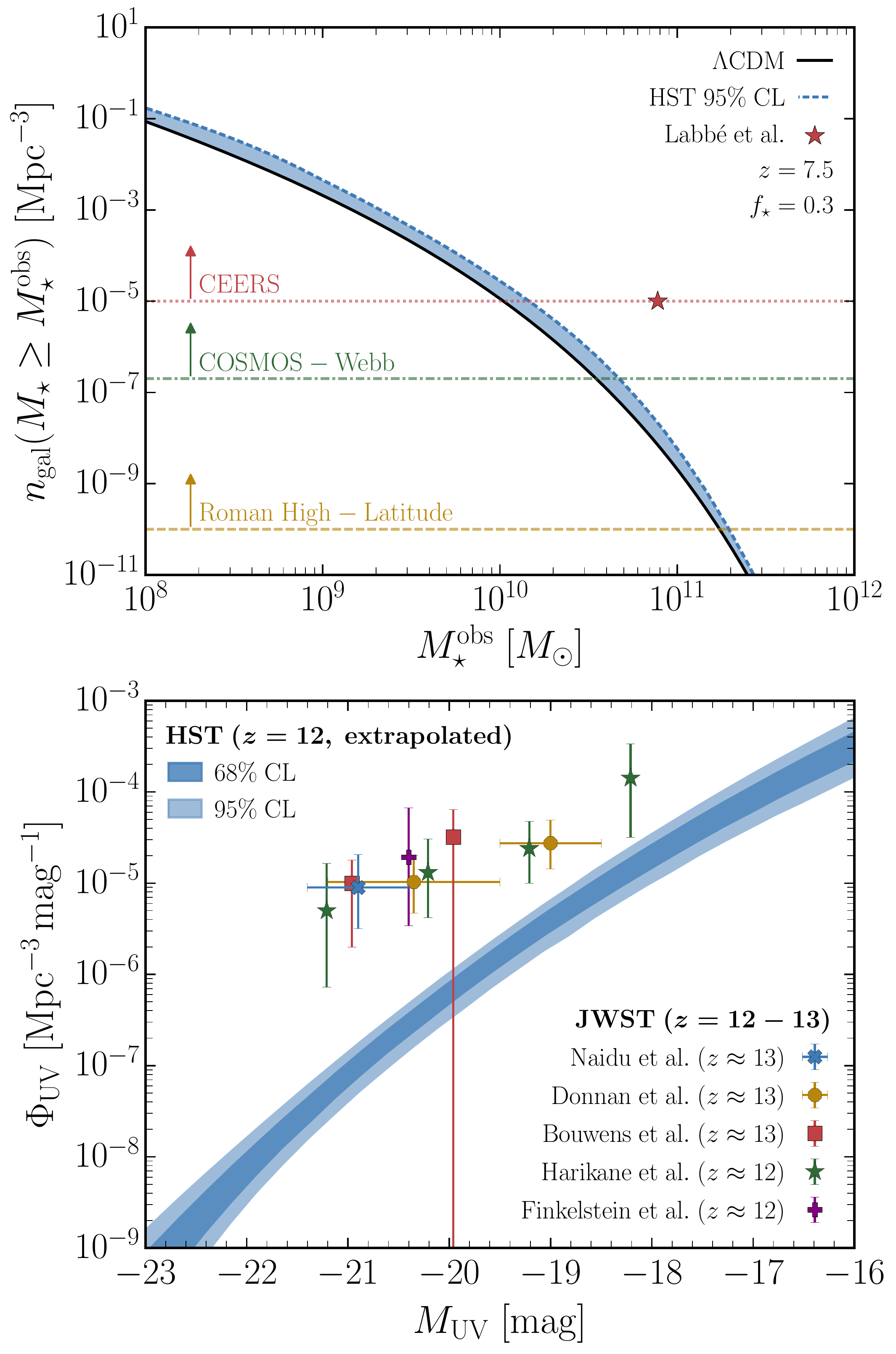}
    \caption{\textbf{Top:} Forecast for the stellar mass function (SMF) at $z=7.5$, defined as the number density of galaxies above a stellar mass cut-off $M_\star^{\rm obs}$. The black curve is the $\Lambda$CDM prediction, and the blue dotted curve indicates the enhancement allowed by HST UV LF data at 95\% CL, all assuming a constant $f_\star=0.3$ (which can be rescaled to any $\tilde f_\star$ by simply shifting the curves horizontally by a factor $\tilde f_\star/0.3$). Horizontal lines represent the fiducial volume (i.e., 1/Vol) of different surveys~\cite{Finkelstein_CEERS_I1,Casey:2022amu,Spergel:2015sza}, whereas the red star is the inferred density from the $z = 7.5$ galaxy of Ref.~\cite{Labbe_2023}, which is far in excess of the SMF allowed by HST. \textbf{Bottom:} UV luminosity function at $z = 12$, extrapolated from HST calibrations at $z = 6 - 10$, along with JWST measurements at $z = 12 - 13$ from~\cite{Naidu_UVLF_2022,2023MNRAS.518.6011D, Bouwens:2022gqg, Harikane_UVLFs,2022ApJ...940L..55F}. The 68\% and 95\% confidence levels are obtained after marginalizing over our astrophysical parameters and the enhancement of power. We find that including a power boost alone is not enough to explain the JWST UV LFs.
    }
    \label{fig:HST_forecast}
\end{figure}

\textit{Forecasts.\ ---}
Future data from JWST and upcoming telescopes like Roman will yield additional measurements of the stellar mass function during reionization and cosmic dawn. Here, we leverage our findings to forecast the maximum expected galaxy number density permitted by the current HST UV LFs. We use the results from the HST MCMC analysis above to calculate the projected density of galaxies at redshift $z=7.5$ as a function of the cut-off stellar mass $M_\star^\mathrm{obs}$, assuming an optimistic star-formation efficiency of $f_\star=0.3$. We show the results in the top panel of Fig.~\ref{fig:HST_forecast}. Note that, unlike in Fig.~\ref{fig:Ngal_distributions}, we do not multiply by a volume element, and show the 95\% confidence limits (CL) allowed by HST instead of the full posterior distributions. For reference, we compare these projected densities with the typical volumes probed by the CEERS JWST data ($\sim10^5$ Mpc$^3$~\cite{Finkelstein_CEERS_I1}), the forthcoming COSMOS-Webb survey ($\sim5\times 10^6$ Mpc$^3~$\cite{Casey:2022amu}), and a Roman high-latitude survey ($\sim10^{10}$ Mpc$^3$~\cite{Spergel:2015sza}). Our analysis reveals that the HST UV LFs only allow for a modest increase in the number density of galaxies at this redshift. The figure includes the inferred density from the $z = 7.5$ JWST galaxy candidate of Ref.~\cite{Labbe_2023} (ID 38094), which significantly exceeds our 95\% CL region. Interestingly, we find that even in the extreme scenario of $f_\star = 1$ the JWST point still lies above our 95\% CL curve, implying a severe inconsistency with the HST UV LFs. 
A similar statement was recently made in Ref.~\cite{Prada:2023dix} within the context of $\Lambda$CDM-only simulations (black line in Fig.~\ref{fig:HST_forecast}), and in Ref.~\cite{Parashari:2023cui} for cosmologies with blue-tilted power spectra.

We note that the issue of ultra-massive galaxy candidates at $z = 7-10$ is distinct from a possible excess in the UV LF at $z>10$. The latter arises from galaxies seemingly boasting star-formation rates far above our predicted values at earlier times~\cite{Gnz11_Oesch,Yung:2023bng,Dekel:2023ddd,McCaffrey_notension_z10,Kannan:2022szh}. While Refs.~\cite{Mason:2022tiy,Haslbauer:2022vnq, Shen:2023cva, 2023arXiv230715305S} argued that this excess does not necessarily challenge the $\Lambda$CDM cosmological model, as large UV brightness may arise if star formation occurs very rapidly or there is a sizeable UV variability, there is still a discrepancy between the UV LFs measured by JWST at $z>10$ and predictions calibrated with HST data at lower redshifts~\cite{Mirocha_UVLFs2023}. We showcase this discrepancy in the bottom panel of Fig.~\ref{fig:HST_forecast}, where we employ the same HST analysis in \texttt{GALLUMI} from before to demonstrate that any enhancement of power allowed by HST observations (at $z=6-10$) cannot account for the excessive UV brightness observed in JWST at $z>10$. This hints at an astrophysical, rather than cosmological, explanation of the UV-bright galaxies beyond $z = 10$, see also Ref.~\cite{Munoz:2023cup}.\\

\textit{Conclusion.\ ---}
The emergence of JWST data has opened a new avenue to understand the first galaxies and their cosmological context. The first JWST observations have possibly revealed an excess of ultra-massive galaxies, challenging our standard $\Lambda$CDM cosmological model. In this work, we have shown that contemporaneous measurements of the UV galaxy luminosity function by the Hubble Space Telescope rule out the required departures from $\Lambda$CDM, disfavoring a cosmological solution to the JWST abundance problem. Moving forward, we argue that leveraging the well-established knowledge from HST data will be key for interpreting future observations of the first galaxies by JWST and the upcoming Roman Space Telescope.\\

\textit{Acknowledgements.\ ---} We are very thankful to M. S. Turner for discussions that sparked this work, and to M. Boylan-Kolchin, S. Finkelstein, C. Mason, and J. Mirocha for their insightful comments on a previous version of this manuscript. NS was supported by a Horizon Fellowship from Johns Hopkins University. JBM was supported by NSF grant No.\ 2307354. MK was supported by NSF Grant No.\ 1818899 and the Simons Foundation.

\bibliographystyle{apsrev4-1}
\bibliography{biblio}

\clearpage
\newpage

\onecolumngrid
\section*{Supplementary Material}

\section*{Posteriors from Power Boost}

\begin{figure}[h!]
    \centering
    \includegraphics[width=0.55\linewidth]{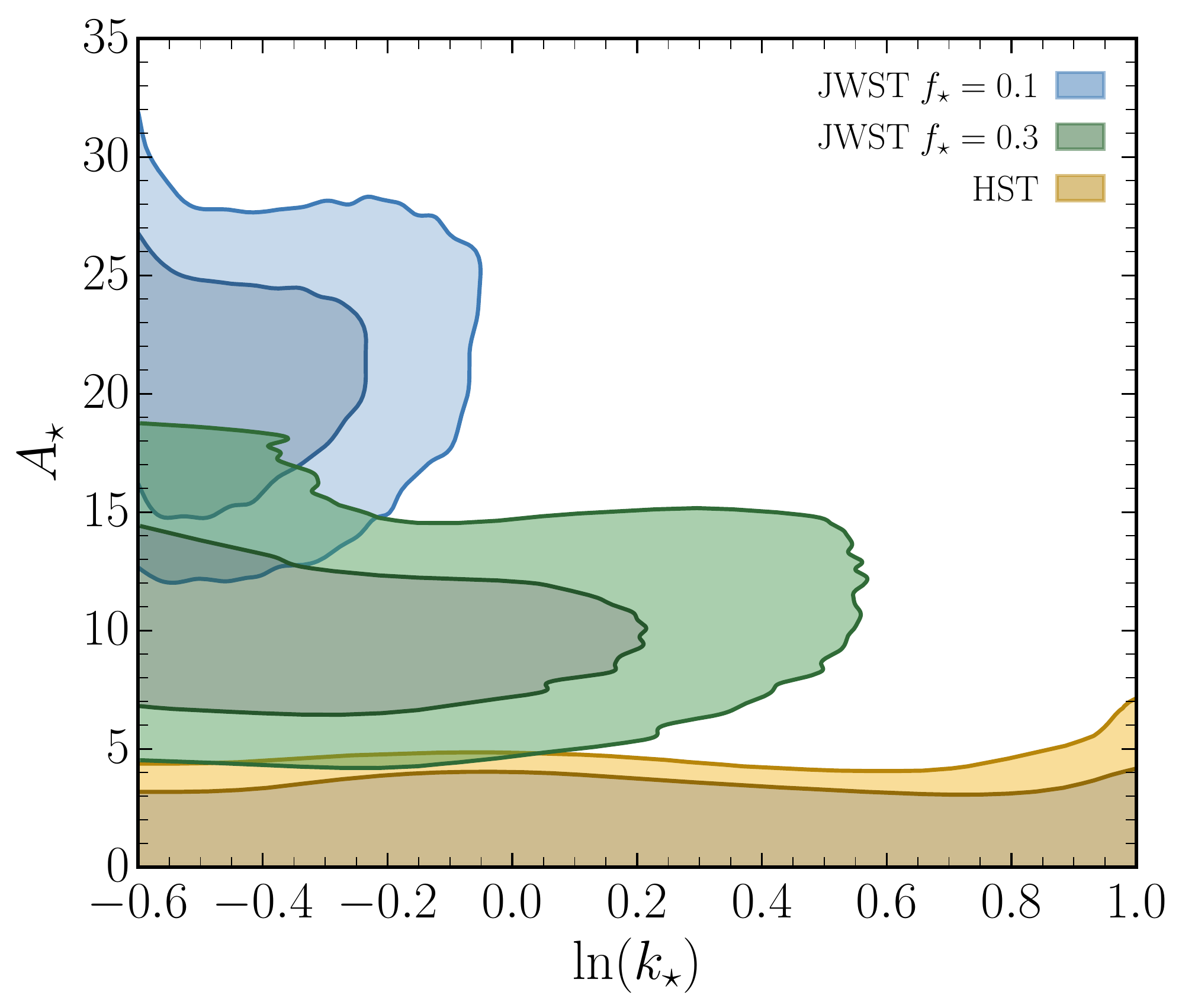}
    \caption{Posteriors for the amplitude and mean of the power boost in Eq.~\eqref{eq:bump}, with $\sigma_\star = 0.1$. The blue and green posteriors are obtained in the JWST analysis using $f_\star = 0.1$ and $f_\star = 0.3$, respectively, while the yellow one is based on the HST analysis. The inner (outer) contours depict the 68\% (95\%) confidence levels.
    }
    \label{fig:power_2D_posteriors}
\end{figure}

In Fig.~\ref{fig:power_2D_posteriors}, we show the 2D posteriors on the amplitude and mean of the power boost as parameterized in Eq.~\eqref{eq:bump}. The blue and green contours show which values of the amplitude and mean are required to account for the abundance of the JWST galaxy candidates, using a star-formation efficiency of $f_\star = 0.1$ and $f_\star = 0.3$ respectively, while the yellow one shows how these same quantities are constrained by the Hubble UV galaxy luminosity function. It is clear that the JWST posteriors do not overlap with the HST one, highlighting our main result of this work that an enhancement of the matter power spectrum cannot account for the abundance problem.

\section*{Robustness Checks}
Throughout this paper, we have followed the approach of Refs.~\cite{Sabti:2021unj,Sabti:2021xvh} to model the UV LFs from HST. Here, we will perform cross-checks of the key assumptions in the modeling, and study how they affect our conclusions. Before that, we note that our MCMC simulations in MontePython are done using the default settings and run twice for each scenario considered in this work, such that the second run does not have a burn-in phase and convergence can be achieved quicker. The simulations are stopped once the posteriors are smooth enough and do not change shape. We find that changing the shape of power enhancenment (either by adding a broad-band component, or altering the bump width), the halo mass function (HMF), or the dust calibration does not alter our conclusions, as summarized in Fig.~\ref{fig:robustness_checks}.
We encourage the reader to visit Ref.~\cite{Sabti:2021xvh} for full corner plots with all astrophysical parameters and their correlations with cosmology.

\subsection*{Broad-band Enhancement of Power}

Thus far, we have assumed that any enhancement of power is limited to scales $k\geq 0.5\,\Mpcinv$, as the matter power spectrum has been accurately measured at lower values of $k$ by other cosmic data sets. Nevertheless, it is possible to construct a scenario where the redshift range that has been well-measured ($z\sim0$ and $z\sim 10^3$) is consistent with the $\Lambda$CDM model, while during the epoch of reionization the power is overall higher, even at large scales (low $k$). This is generically the prediction of models that change the growth factor~\cite{Menci:2022wia}. Although this may be inconsistent with existing CMB-lensing measurements (as their kernel covers these $z$), here we will show that it is also disfavored by the HST UV LFs. For this purpose, we will vary the amplitude $\sigma_8$ of fluctuations in our analysis, along with the bump $A(k)$ from Eq.~\eqref{eq:bump}. We find that even with the extra degree of freedom, the UV LF data do not allow for a substantial increase in the number of galaxies. This test guarantees that a broadband enhancement of power during reionization ($z=6-10$) is insufficient to explain the JWST galaxies from Ref.~\cite{Labbe_2023}, as shown in Fig.~\ref{fig:robustness_checks}. As an additional check, we have also considered a case where on top of the bump $A(k)$ in Eq.~\eqref{eq:bump} all $\Lambda$CDM cosmological parameters are varied within the Planck limits~\cite{Planck:2018vyg} (by adding Gaussian priors to the likelihood), and again found no discernible difference.

\subsection*{Halo Mass Function}

Here we test how our results depend on the assumed halo mass function at high $z$. In all cases, we have used the Sheth-Tormen fit, where the collapsed fraction is~\cite{Sheth:2001dp}:
\begin{align}
    f_\mathrm{ST}(\nu) = A_\mathrm{ST} \sqrt{\dfrac{2}{\pi}}\left [ 1 + \nu^{-2 p_\mathrm{ST}} \right] \nu \exp ( -\nu^2/2)\ ,
\end{align}
with $\nu = \sqrt{a_\mathrm{ST}} \delta_{\rm crit}/\sigma$. Here, $\delta_{\rm crit} = 1.686$ is the threshold for collapse, $\sigma^2$ is the variance of fluctuations, and the $(\ldots)_\mathrm{ST}$ parameters are fitted to simulation results. In this work, we used the fit from Ref.~\cite{Schneider:2020xmf}, with $a_\mathrm{ST} = 0.85$, $p_\mathrm{ST}=0.3$, and  $A_\mathrm{ST}=0.3222$. We will repeat our analysis with the Reed mass function fit~\cite{Reed:2006rw}, which is given by:
\begin{align}
\label{eq:Reed_HMF}
    f_\mathrm{Rd}(\nu) = & A_\mathrm{Rd}\sqrt{\frac{2}{\pi}}\left[1+\nu^{-2 p_\mathrm{Rd}} + 0.2\exp\left(-\frac{\left(\ln\sigma^{-1}-0.4\right)^2}{2(0.6)^2}\right)\right]\nu\exp\left(-c_\mathrm{Rd}\nu^2/2\right)\ ,
\end{align}
where $\nu = \sqrt{a_\mathrm{Rd}} \delta_{\rm crit}/\sigma$, $A_\mathrm{Rd} = 0.3235$, $a_\mathrm{Rd} = 0.707$, $p_\mathrm{Rd} = 0.3$, and $c_\mathrm{Rd} = 1.081$. 
The Reed mass function tends to predict a higher halo abundance at high redshifts than the Sheth-Tormen calibration, leading to a larger number of galaxies (in both the UV and stellar mass functions), though still not sufficient to reach the reported abundance of ultra-massive galaxies, as can be seen in Fig.~\ref{fig:robustness_checks}.

\subsection*{Bump Shape}

In our analysis, we made the assumption that the boost $A(k)$ in the matter power spectrum has a fixed width $\sigma_\star=0.1$, but a variable amplitude $A_\star$ and location $k_\star$. This was sufficient to capture variations of the HMF that can reproduce the abundance of JWST galaxies (as demonstrated by the JWST results in Fig.~\ref{fig:Ngal_distributions}). Here we relax this assumption and allow the width $\sigma_\star$ to vary in the analysis. As expected, this adjustment does not result in an increase in the number of galaxies, see Fig.~\ref{fig:robustness_checks}.

\subsection*{Dust Attenuation}

\begin{figure}[t!]
    \centering
    \includegraphics[width=0.55\linewidth]{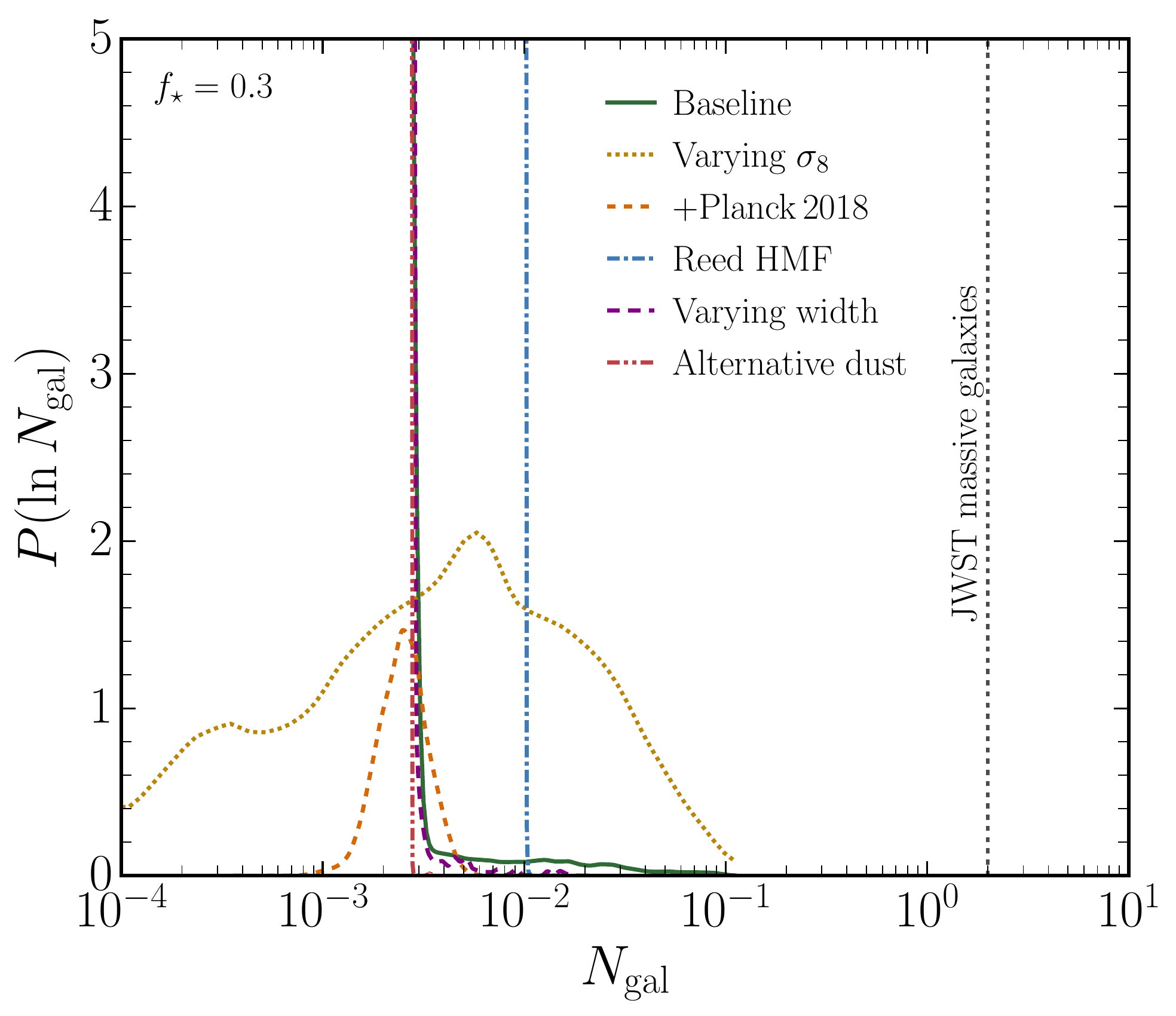}
    \caption{Probability distributions of the average number of galaxies, as in Fig.~\ref{fig:Ngal_distributions}, assuming a constant $f_\star=0.3$.
    We show our baseline case as described in the main text (green solid), a case with a broad-band enhancement of power through $\sigma_8$ (yellow dotted), one where the $\Lambda$CDM cosmological parameters are also varied but confined within Planck 2018 limits~\cite{Planck:2018vyg} (orange dashed), a case with a different halo mass function (blue dash-dotted), one where the width $\sigma_\star$ of the boost is varied (purple dashed), and, finally, one where a different dust calibration is used (red double dash-dotted). None of these alterations reaches $N_{\rm gal}=2$ (black dotted), disfavoring a cosmological explanation to the abundance of ultra-massive galaxies reported in Ref.~\cite{Labbe_2023}.
    }
    \label{fig:robustness_checks}
\end{figure}

Dust extinction is known to affect the UV LFs, on average attenuating the bright end. While this is a bigger issue at lower $z$~\cite{Yung_2018}, here we will examine how much it can change our results at $z=6-10$. We use the IRX-$\beta$ relation from~\cite{Meurer:1999jj}, where the attenuation is given by:
\begin{align}
\langle A_\mathrm{UV}\rangle = C_0 + 0.2\ln(10)\sigma_\beta^2 C_1^2+ C_1\langle\beta\rangle\ ,
\end{align}
with the $C_i$ free parameters that are fit to observations, and $\sigma_\beta = 0.34$~\cite{Bouwens:2011yy}. The quantity $\langle \beta\rangle$ represents the average power-law index of the UV spectrum and is modeled using observations from~\cite{Bouwens:2013hxa}, see Ref.~\cite{Sabti:2021xvh} for more details. In the main text we adopted the $C_i$ from~\cite{Overzier:2010aa}, with $C_0 = 4.54$ and $C_1 = 2.07$, whereas for this test we will use a different calibration from~\cite{Casey:2014cqa}, with $C_0 = 3.36$ and $C_1 = 2.04$. The latter calibration results in a weaker dust attenuation, bringing the observed UV LFs closer to the intrinsic ones, which in turn implies fewer bright galaxies. As such, we would expect less freedom in enhancing the number of galaxies, as we confirm in our analysis in Fig.~\ref{fig:robustness_checks}. 

\end{document}